\documentclass[twocolumn,showpacs,preprintnumbers,amsmath,amssymb]{revtex4}

\usepackage{graphicx}
\usepackage{float}
\usepackage{dcolumn}
\usepackage{bm}

\begin{document}

\title{Effect of Co substitution on Ni$_{2}$MnGe Heusler alloy: \textit{ab initio}  study}

\author{M. Pugaczowa-Michalska\footnote{corresponding author e-mail:maria@ifmpan.poznan.pl}}
\affiliation{%
Institute of Molecular Physics, Polish Academy of Sciences\\M. Smoluchowskiego 17,
60-179 Pozna\'{n}, Poland
}%
\date{\today}

\begin{abstract}
\textit{Ab initio} calculations shown that  the Co substitution instead of Ni in Ni$_{2}$MnGe with the L2$_{1}$ crystallographic structure leads to a decrease of the lattice constant and an increase of the total magnetic moment of the Ni$_{2-x}$Co$_{x}$MnGe compounds. The Mn(B) has the largest local moment above $3~\mu_{B}$ coupled parallel to moments on the Ni(A,C) and Co(A,C), which are found in the ranges of $0.19\div0.26~\mu_{B}$ for Ni(A,C) and $1.03\div0.97~\mu_{B}$ for Co(A,C) for studied range of $x$.  Using the results stemming from the total energy calculations, the values of bulk modulus and its pressure derivatives are estimated according to the Murnaghan EOS.
\end{abstract}

\maketitle

\section{Introduction}
Heusler alloys have emerged as one of interesting topic in materials science, triggering a multitude of research activities with a focus on both fundamental science as well as tailored applications. Partial or complete substitution of atoms by various elements or changing of relative concentrations of atoms in well known Heusler alloys can help in tuning of their various properties \cite{Chakrab, Roy}. The magnetic properties of  Ni$_{2-x}$Co$_{x}$MnGe ($0\leq x\leq 1.0$) system should be interesting because from first principles calculations the end compounds  Ni$_2$MnGe and Co$_{2}$MnGe provide two different possibilities of application. The Co$_{2}$MnGe was predicted to be half metals if ferromagnetic states are assumed \cite{Ishida1}. However, the other end compound Ni$_2$MnGe in the full ordered L2$_1$ structure is a typical metal \cite{MPM1}. The properties of ferromagnetic ground state of Ni$_{2}$MnGe Heusler alloy as well as the magnetic properies under hydrostatic pressure have been reported in \cite{MPM1}. The studies of Ni or Fe substitution at B site in Ni$_{2}$MnGe in the framework of the coherent potential approximation or the virtual crystal approximation demonstrated that the total spin magnetic moment decreases with increase of Ni or Fe concentration \cite{MPM3, Wei}. However, the magnetic properties of Ni$_2$MnGa$_{1-x}$Ge$_{x}$ have remained basically unchanged due to substitution at D site \cite{Kaczkowski}. 

The experimental evidence for Ni$_2$MnGe is that the site disorder can hardly be avoided completely in epitaxial Ni$_2$MnGe film grown by molecular beam epitaxy \cite{Lund}. Perfect site order in the Heusler unit cell is difficult to achieve even in bulk single crystals. Thus, some of attempts to obtain samples with ordered structure are directed to substitution of one element by another atoms.

One of motivations for studying the Ni$_{2-x}$Co$_{x}$MnGe ($0 \leq x\leq 2.0$) is to search how A site substitution influences to electronic properties of the systems. The effect of Co-substitution on magnetic properties of Ni$_2$MnGe will be discussed at \textit{ab initio} level using results of the FPLO-CPA calculations. 

\section{Computational tools}
The numerical calculations were done using the full-potetial nonorthogonal local orbital basis (FPLO) scheme \cite{Koepernik1, Opahle} within the local spin density approximation (LSDA). In the scalar-relativistic calculations the exchange and correlation potential of the Perdew and Wang was chosen \cite{Perdew}. The atomic disorder at the specific site was treated as the Coherent Phase Approximation (CPA), covered by the CPA solver described in \cite{Koepernik2}. 
The unit cell of the studied Ni$_{2-x}$Co$_{x}$MnGe ($0 \leq x \leq 2.0$) system is the cubic L2$_{1}$, which contains four interpenetrating {\it fcc} sublattices at A(0~0~0), B($\frac{1}{4}$ $\frac{1}{4}$ $\frac{1}{4}$), C($\frac{1}{2}$ $\frac{1}{2}$ $\frac{1}{2}$) and D($\frac{3}{4}$ $\frac{3}{4}$ $\frac{3}{4}$) sites. In the end compound Ni$_{2}$MnGe, these sites are occupied by Ni, Mn, Ni and Ge atoms, respectively. In present calculations, the A and C positions are occupied both by Ni and Co atoms. The following concentrations of Co atom are considered: $x=0, 0.2, 0.4, 0.6, 0.8, 1.0, 1.2, 1.4, 1.6, 1.8$. The Brillouin zone (BZ) is sampled by \textit {k}-mesh containing 256 points in its irreducible part. The other details of the FPLO calculation were similiar to \cite{MPM3}.

\section{Results}
\begin{figure*}
\centering
\includegraphics[width=0.88\textwidth]{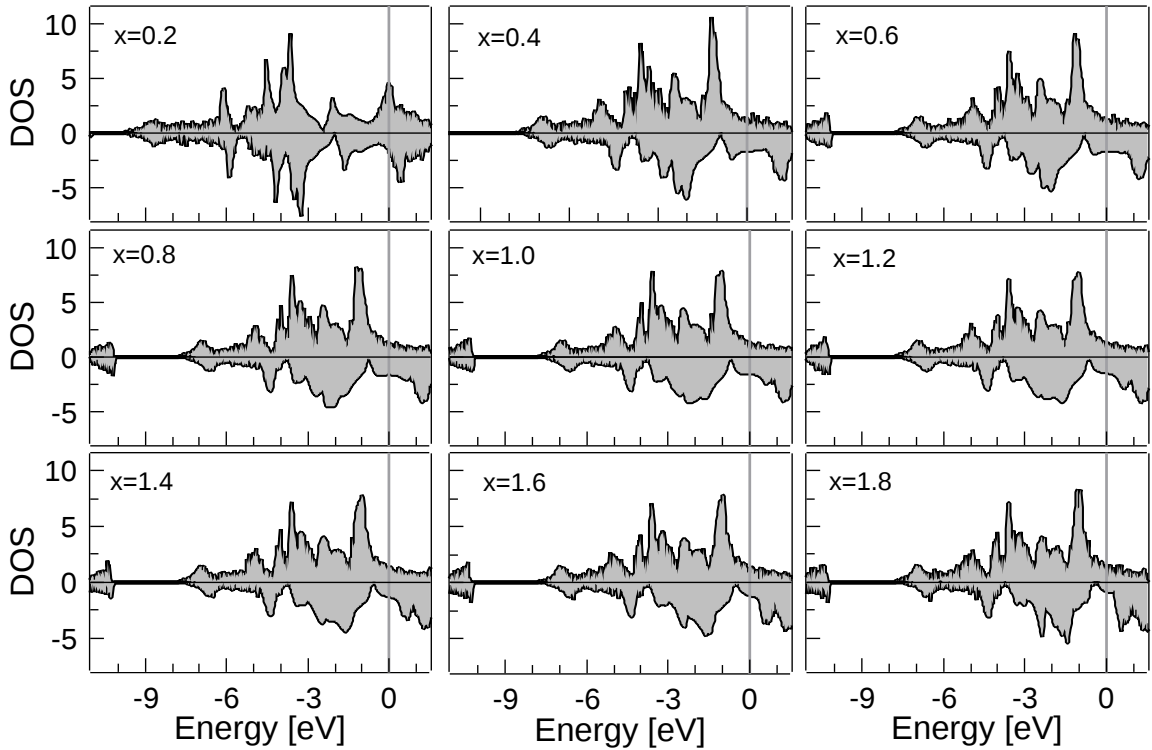}
\caption{\label{fig:1} Total density of states of Ni$_{2-x}$Co$_x$MnGe. DOS is in [states/eV]. The vertical line is the Fermi level ($E_{F}$=0).}
\end{figure*}

We studied electronic structure of  Ni$_{2-x}$Co$_x$MnGe assuming that Co atoms are simultaneously located at A and C site in the unit cell. The decrease in lattice constant of Ni$_2$MnGe with Co-substitution was obtained (Table 1). 

The total electronic densities of states (DOS) of the studied system (Fig.\ref{fig:1}) show that the reconstruction of the spin dependent bands with Co substitutionis mainly driven by hybridisation. The smearing of some peaks in the total DOS for spin-down direction with an increase of Co concentration is observed. The high peak of the DOS for spin-up direction in proximity of the Fermi energy ($E_{F}$) was obtained for Ni$_{1.8}$Co$_{0.2}$MnGe. However, for higher Co concentration ($x\ge0.4$) this peak is shifted below the Fermi level. The partial filling of some valleys between main peaks in the total DOS for spin-up direction are visible with an increase of Co concentration. 
\begin{figure}[H]
 \includegraphics[scale=0.48]{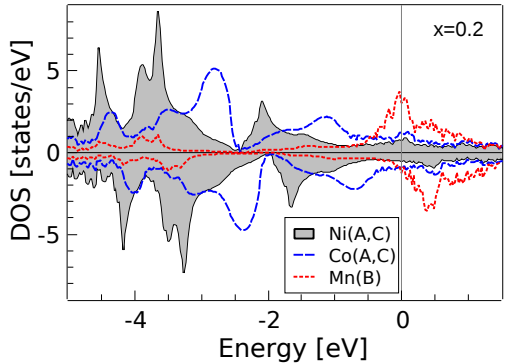}
 \caption{\label{fig:2} The atom-resolved DOS in Ni$_{1.8}$Co$_{0.2}$MnGe. The vertical line is the Fermi level ($E_{F}$=0).}
\end{figure}
We now turn our attention to the peak at $E_{F}$ in the DOS of Ni$_{1.8}$Co$_{0.2}$MnGe. The structural instability of the cubic Heusler phase is typically indicated by such van Hove singularities at the Fermi level. One can expect that reasonable electronic relaxation mechanism, or magnetism, or a structural distortion reduce the DOS at $E_{F}$ in Ni$_{1.8}$Co$_{0.2}$MnGe (i.e. eliminate this singularity). Detaled analysis of the atomic resolved DOS of Ni$_{1.8}$Co$_{0.2}$MnGe (Fig.\ref{fig:2}) reveals that the contribution to the peak in the total DOS at $E_{F}$ comes mainly from Mn(B). The atom-resolved DOS provide evidence that electronic states of \textit{3d} Ni and \textit{3d} Co contribute to these peaks, also. However, it is clear that  the overall trend in majority DOS at $E_{F}$ for the system with $x=0.2$ is dictated by \textit{3d} states of Mn(B). The unoccupied DOS at $E_{F}$ and above derives mainly from the minority bands of Mn(B). 

\begin{figure}
 \includegraphics[scale=0.48]{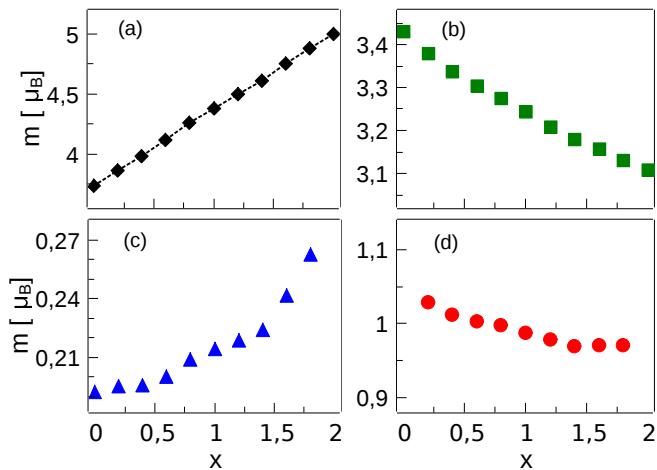}
 \caption{\label{fig:3} Total magnetic moment (a) and contributions associated with individual atoms of Mn (b), Ni(c) and Co (d) in Ni$_{2-x}$Co$_x$MnGe as a  function of concentration $x$}
\end{figure}

Every additional Co in A or C sites, as the first nearest neighbours of Mn(B), has effected to magnetic properties of Mn atom as well as other atoms. By increasing the concentration of Co atoms, the concentration of valence electrons decreases in the studied Ni$_{2-x}$Co$_x$MnGe, because Co has one electron less than Ni in its atomic orbitals. The linear growth of the total magnetic moment obtained in Ni$_{2-x}$Co$_x$MnGe (Fig.\ref{fig:3}) is in accordance with a Slater-Pauling type curve for the total spin-magnetic moment for binary alloys Ni-Co \cite{Bozorth}. Most of the magnetic properties of the Heusler ferromagnetic alloys, like Ni$_2$MnGe or Co$_2$MnGe, stem from the magnetic moments of the Mn atoms. In the studied Ni$_{2-x}$Co$_x$MnGe the Mn atoms are far from each other in the the cubic L2$_{1}$ structure, i.e. third nearest neighbor (NN) of each other regardless from Co concentration, which implies an indirect exchange mechanism. The calculated spin magnetic moments at Mn(B) are above $3~\mu_B$ in studied Ni$_{2-x}$Co$_x$MnGe for the range of studied Co concetration. The values of Mn moment slightly decrease with $x$ from $3.38~\mu_B$ for $x=0.2$ to $3.13~\mu_B$ for $x=1.8$. However, by Co substitution, the total magnetic moment increases which is due to the magnetic moment of Ni and Co atoms. The contributions of Ni to the magnetic moment modulates the overall trend dictated by Mn (due to its being largest), hence the total moment reflects a similar trend to that of Ni. The values of Ni moment obtained in our calculations ranges from $0.192~\mu_B$ for $x=0$ to $0.262~\mu_B$ for $x = 1.8$. The Co moment changes from $1.028~\mu_B$ for $x=0.2$ to $0.972~\mu_B$ for $x=2.0$. A small negative magnetic moment ($-0.08~\mu_B \div -0.05~\mu_B$) is also induced on Ge atom in the studied alloys. The calculated highest value of total magnetic moment of $4.999~\mu_B$ is achieved for Co$_2$MnGe.   

Structural properties of the studied Heusler compounds are computed by energy minimization procedure. The calculated total energy of the Ni$_{2-x}$Co$_x$MnGe as a function of unit cell volume ($V$) is fitted according to the Murnaghan isothermal equation of states (EOS) \cite{Murnaghan}:

\begin{equation}
 \label{eq:1} 
    E_{tot}(V)=E_{0}+ \frac{B_{0}V}{B^{'}(B^{'}-1)}\Big [ \left(\frac{V_{0}}{V} \right)^{B^{'}}+B^{'}\left(1-\frac{V_{0}}{V}\right)-1\Big ],                                       
\end{equation}
where $B_0$ and $B^{'}$ are the bulk modulus and its pressure derivative at equilibrium volume $V_0$. The computed structural parameters of  Ni$_{2-x}$Co$_x$MnGe are given in Table 1.
\begin{table}[H]
\caption{The calculated theoretical lattice parameter $a_{th}$ corresponding to the minimum of the total energy, bulk modulus and its pressure derivatives for Ni$_{2-x}$Co$_x$MnGe}

\begin{tabular}{ c r l l l l l l} \hline
     &  $a_{th}$ (\AA) & & $B_{0}$ (GPa) & & $B^{'}$ & \\
\hline
  Ni$_{1.8}$Co$_{0.2}$MnGe    &  $5.665$   & &  $203 \pm 2$ & & $1.6$ & \\
  
  Ni$_{1.6}$Co$_{0.4}$MnGe    &  $5.659$   & &  $198 \pm 2$ & & $2.0$ & \\
 
  Ni$_{1.4}$Co$_{0.6}$MnGe     &  $5.654$   & &  $199 \pm 2$ & & $2.5$ & \\
 
  Ni$_{1.2}$Co$_{0.8}$MnGe     &  $5.648$   & &  $208 \pm 1$ & & $4.5$ & \\
  
  Ni$_{1.0}$Co$_{1.0}$MnGe     &  $5.643$   & &  $216 \pm 1$ & & $3.5$ & \\
  
  Ni$_{0.8}$Co$_{1.2}$MnGe     &  $5.637$   & &  $218 \pm 1$ & & $3.2$ & \\
  
  Ni$_{0.6}$Co$_{1.4}$MnGe     &  $5.631$   & &  $225 \pm 1$ & & $3.7$ & \\
  
  Ni$_{0.4}$Co$_{1.6}$MnGe     &  $5.625$   & &  $230 \pm 1$ & & $2.4$ & \\
  
  Ni$_{0.2}$Co$_{1.8}$MnGe     &  $5.621$   & &  $233 \pm 1$ & & $3.0$ & \\
\hline
\end{tabular}
\end{table}
\section{Conclusions}
To summarize, \textit{ab initio} studies show that the Co substitution instead of Ni in parent Ni$_{2}$MnGe does not change magnetic ordering: the Ni$_{2-x}$Co$_{x}$MnGe remains ferromagnetically ordered. However, the total magnetic moment versus concentration $x$ for Ni$_{2-x}$Co$_{x}$MnGe
reveals an almost linear relationship. The electronic structure calculations of Ni$_{1.8}$Co$_{0.2}$MnGe exhibit a marked peaks in DOS at the Fermi level that may consider as a source of possible instability of the system. The calculated DOS for other compounds from series Ni$_{2-x}$Co$_{x}$MnGe with higher Co content have not any singularity in the DOS at $E_{F}$. Our study provides the values of bulk modulus and its pressure derivatives according to the isothermal EOS.

\def\refname{References}


\begin{thebibliography}{20}
\bibitem{Chakrab} A. Chakrabarti, M. Siewert, T. Roy, K. Mondal, A. Banerjee, Markus E. Gruner, and P. Entel, {\em Phys. Rev. B} {\bf 88}174116 (2013).
\bibitem{Roy} T. Roy, Markus E. Gruner, P. Entel, A. Chakrabarti, {\em J. Alloys Compnd.} {\bf 632} 822 (2015).
\bibitem{Ishida1} S. Ishida, T. Masaki, S. Fujii, S. Asano, {\em Physica B: Cond. Mater.} {\bf 245} 1 (1998);  {\em J. Phys. Soc. Jpn.} {\bf 64} 2152 (1995). 
\bibitem{MPM1} M. Pugaczowa-Michalska, {\em J. Alloys Compnd.} {\bf 427} 54 (2007). doi:10.1016/j.jallcom.2006.03.036
\bibitem{MPM3} M. Pugaczowa-Michalska, {\em Comput.  Mater.  Sci.} {\bf  50} 15 (2010). doi:10.1016/j.commatsci.2010.07.002
\bibitem{Wei} X.P. Wei, Y.D. Chu, X.W. Sun, J.B. Deng, Y. Z. Xing, {\em J. Supercond. Nov. Magn.} {\bf 27} (2014) 1099-1103. 	DOI:	10.1007/s10948-013-2370-6
\bibitem{Kaczkowski} M. Pugaczowa-Michalska, A. Jezierski, J. Dubowik, J. Kaczkowski, {\em Acta Phys. Polon. A} {\bf 115} 241 (2009).
\bibitem{Lund}  M.S. Lund, J.W. Dong, J. Lu, X.Y. Dong, C.J. Palmstr{\o}m, C. Leighton, {\em Appl. Phys. Lett.} {\bf 80} 4798 (2002).
\bibitem{Koepernik1} K. Koepernik, H. Eschrig, {\em Phys. Rev. B} {\bf 59} 1743 (1999).
\bibitem{Opahle} I. Opahle, K. Koepernik, H. Eschrig, {\em Phys. Rev. B} {\bf 60} 14035 (1999).
\bibitem{Perdew} J.P.  Perdew, Y. Wang, {\em Phys. Rev. B} {\bf 45} 13244 (1992).
\bibitem{Koepernik2} K. Koepernik, B. Velicky, R. Hayn, H. Eschrig, {\em Phys. Rev. B}  {\bf 55} 5717 (1997).
\bibitem{Bozorth} R.M. Bozorth, {\em Ferromagnetism}, (1951) D. Van Nostrang Company, New York.
\bibitem{Murnaghan} F.D. Murnaghan, {\em Proc. Natl.Acad. Sci USA} {\bf 30} (1944) 244.
\end{thebibliography}
\end{document}